\documentclass[12pt,preprint]{aastex61}

\usepackage{hyperref}
\usepackage{cleveref}

\newcommand\aastex{AAS\TeX}

\received{October 26, 2017}
\revised{November 16, 2016}
\accepted{\today}
\submitjournal{ApJ}

\shorttitle{\aastex\ Stellar parameters for TRAPPIST-1}
\shortauthors{Van Grootel et al.}

\begin{document}

\title{Stellar parameters for TRAPPIST-1}

\correspondingauthor{Val\'erie Van Grootel}
\email{valerie.vangrootel@uliege.be}

\author{Val\'erie Van Grootel}
\affil{Space sciences, Technologies and Astrophysics Research (STAR) Institute, Universit\' e
  de Li\` ege, 19C All\'ee du 6 Ao\^ ut, B-4000 Li\` ege, Belgium}

\author{Catarina S. Fernandes}
\affiliation{Space sciences, Technologies and Astrophysics Research (STAR) Institute, Universit\' e
  de Li\` ege, 19C All\'ee du 6 Ao\^ ut, B-4000 Li\` ege, Belgium}

\author{Michael Gillon}
\affiliation{Space sciences, Technologies and Astrophysics Research (STAR) Institute, Universit\' e
  de Li\` ege, 19C All\'ee du 6 Ao\^ ut, B-4000 Li\` ege, Belgium}
\nocollaboration

\author{Emmanuel Jehin}
\affiliation{Space sciences, Technologies and Astrophysics Research (STAR) Institute, Universit\' e
  de Li\` ege, 19C All\'ee du 6 Ao\^ ut, B-4000 Li\` ege, Belgium}

\author{Jean Manfroid}
\affiliation{Space sciences, Technologies and Astrophysics Research (STAR) Institute, Universit\' e
  de Li\` ege, 19C All\'ee du 6 Ao\^ ut, B-4000 Li\` ege, Belgium}

\author{Richard Scuflaire}
\affiliation{Space sciences, Technologies and Astrophysics Research (STAR) Institute, Universit\' e
  de Li\` ege, 19C All\'ee du 6 Ao\^ ut, B-4000 Li\` ege, Belgium}
  
\author{Adam J. Burgasser}
\affiliation{Center for Astrophysics and Space Science, University of California San Diego, La Jolla, CA, 92093, USA}  
\affiliation{Fulbright Scholar, University of Exeter, College of Engineering, Mathematics and Physical Sciences, Exeter, EX4
4QL, UK}  

\author{Artem Burdanov}
\affiliation{Space sciences, Technologies and Astrophysics Research (STAR) Institute, Universit\' e
  de Li\` ege, 19C All\'ee du 6 Ao\^ ut, B-4000 Li\` ege, Belgium}
  
\author{Laetitia Delrez}
\affiliation{Cavendish Laboratory, J.J. Thomson Avenue, Cambridge, CB3 0HE, UK}

\author{Brice-Olivier Demory}
\affiliation{University of Bern, Center for Space and Habitability, Gesellschaftsstrasse 6, CH-3012, Bern, Switzerland}

\author{Julien de Wit}
\affiliation{Department of Earth, Atmospheric and Planetary Sciences, Massachusetts Institute of Technology, 77 Massachusetts
Avenue, Cambridge, MA 02139, USA}

\author{Didier Queloz}
\affiliation{Cavendish Laboratory, J.J. Thomson Avenue, Cambridge, CB3 0HE, UK}

\author{Amaury H.M.J. Triaud}
\affiliation{School of Physics \& Astronomy, University of Birmingham, Edgbaston, Birmingham B15 2TT, UK}

\begin{abstract}

TRAPPIST-1 is an ultracool dwarf star transited by seven Earth-sized planets, for which thorough characterization of atmospheric properties, surface conditions encompassing habitability and internal compositions is possible with current and next generation telescopes. Accurate modeling of the star is essential to achieve this goal. We aim to obtain updated stellar parameters for TRAPPIST- 1 based on new measurements and evolutionary models, compared to those used in discovery studies. We present a new measurement for the parallax of TRAPPIST-1, 82.4 $\pm$ 0.8 mas, based on 188 epochs of observations with the TRAPPIST and Liverpool Telescopes from 2013 to 2016. This revised parallax yields an updated luminosity of $L_*=(5.22\pm0.19)\times 10^{-4} L_{\odot}$, very close to the previous estimate but almost twice more precise. We next present an updated estimate for TRAPPIST-1 stellar mass, based on two approaches: mass from stellar evolution modeling, and empirical mass derived from dynamical masses of equivalently classified ultracool dwarfs in astrometric binaries. We combine them through a Monte-Carlo approach to derive a semi-empirical estimate for the mass of TRAPPIST-1. We also derive estimate for the radius by combining this mass with stellar density inferred from transits, as well as estimate for the effective temperature from our revised luminosity and radius.  Our final results are $M_*=0.089 \pm 0.006 M_{\odot}$, $R_* = 0.121 \pm 0.003 R_{\odot}$, and $T_{\rm eff} =$ 2516 $\pm$ 41 K. Considering the degree to which TRAPPIST-1 system will be scrutinized in coming years, these revised and more precise stellar parameters should be considered when assessing the properties of TRAPPIST-1 planets.
\end{abstract}


\section{Introduction} \label{sec:intro}

TRAPPIST-1 (2MASS J23062928-0502285) is an ultracool M8 dwarf star located 12 pc from the Sun \citep{2000AJ....120.1085G,2006AJ....132.1234C}. It hosts seven Earth-sized planets, of which three  orbit in the habitable zone \citep{2017Natur.542..456G}. It is the first planetary system found to transit such an extremely low mass, Jupiter-sized star. This favorable planet-to-star ratio opens the possibility to thoroughly characterize the exoplanets, including probing their atmospheric properties, with the current and next-generation telescopes \citep{2016Natur.537...69D,2017A&A...599L...3B,2016MNRAS.461L..92B}. TRAPPIST-1 is a unique system for testing planet formation and evolution theories, and for assessing the prospects for habitability among Earth-sized exoplanets orbiting cool M dwarfs, the most numerous stars in the Galaxy \citep{2010AJ....139.2679B}. Finally, TRAPPIST-1 is a golden target for comparative exoplanetology, by contrasting the atmospheric properties, surface conditions and internal compositions of similar exoplanets orbiting the same star. Determining exoplanetary  properties relies on a detailed knowledge of the host star, notably as observations mostly constraints them relatively to those of the host. In particular, the irradiation of the planets scales as $L_*/a^2$, where $L_*$ is the stellar luminosity and $a$ is the semi major axis of the orbit, which depends on the stellar mass through Kepler's third law \citep{2010ARA&A..48..631S}. The transit depth measures the planet-to-star radius ratio, and hence inference of the planet radius requires knowledge of the stellar radius \citep{2010exop.book...55W}. The transformation of the planetary mass ratios determined by transit timings variations (TTVs) into the planet physical parameters rely on the stellar mass \citep{2005MNRAS.359..567A}. A crucial element in assessing the ability for a planet to retain an atmosphere, therefore its long term habitability, is the time its host star takes to contract onto the main sequence (e.g. Luger \& Barnes 2015). This time is finely sensitive to stellar parameters for very low-mass stars, and contraction time rapidly increases to several gigayears below $\sim$0.10 $M_{\odot}$ (e.g. \citealt{1997A&A...327.1039C}, \citealt{2000ApJ...542..464C}, \citealt{2015A&A...577A..42B} -- hereafter BHAC15; see also our Fig. \ref{f1}).

TRAPPIST-1 is a M8.0$\pm$0.5 star (Gillon et al. 2016; see also, e.g., Gizis et al. 2000, Costa et al. 2006, Burgasser et al. 2015, Burgasser \& Mamajek 2017). Its luminosity ($\log L_*/L_{\odot}=-3.28\pm0.03$ or $L_*=(5.25^{+0.38}_{-0.35})\times 10^{-4} L_{\odot}$) has been determined by \citet{2015ApJ...810..158F} from a nearly complete spectral energy distribution and the parallax measurement of \citet{2006AJ....132.1234C}. The iron abundance of TRAPPIST-1, [Fe/H]$=0.04\pm0.08$ \citep{2016Natur.533..221G}, has been estimated from the calibration of \citet{2014AJ....147..160M}. The prior probability distribution functions (PDFs) that were used by \citet{2016Natur.533..221G,2017Natur.542..456G} for stellar mass, radius, and effective temperature are $M_*=$ 0.082 $\pm$ 0.011 $M_{\odot}$, $R_*=$ 0.114 $\pm$ 0.006 $R_{\odot}$, and $T_{\rm eff} =$ 2555 $\pm$ 85 K. The mass and radius estimates come from evolutionary models, combining estimates from \citet{2015ApJ...810..158F} mainly based on \citet{2000ApJ...542..464C} models and our own estimates based on more recent BHAC15 models (details about this can be found in the "Methods" section of Supplementary Information of \citealt{2016Natur.533..221G}).
The estimate for effective temperature was obtained combining the model radius and the luminosity from \citet{2015ApJ...810..158F}. Filipazzo et al. (2015) constrained the age to be higher than 500 Myr in the absence of sign of youth, but an age of about 500 Myr is actually inferred from \citet{2000ApJ...542..464C} or BHAC15 evolutionary models for a $\sim$ 0.082 $M_{\odot}$ star at the luminosity of Filipazzo (see also our Sect. \ref{sect4} and Fig. \ref{f1}).

However, TRAPPIST-1 is most likely not a young star, as recently argued by \citet{2017ApJ...845..110B} who examined all available age indicators for TRAPPIST-1. Combining age probability distribution functions from metallicity and kinematics, and lower limits from the absence of lithium absorption and measured rotation period, \citet{2017ApJ...845..110B}  inferred an age of 7.6 $\pm$ 2.2 Gyr. The authors also proposed, based on this old age estimate and the observed luminosity, to revise the stellar radius upwards to $R_* = 0.121\pm0.003 R_{\odot}$. They obtained this radius by artificially "inflating" the radii obtained by evolutionary models of \citet{1997ApJ...491..856B,2001RvMP...73..719B} and BHAC15 to account for the stellar density inferred from transits. It is indeed well known that current stellar models tend in many cases to underestimate stellar radii for low-mass stars \citep[e.g.][and references therein]{2005nlds.book.....R,2013ApJ...776...87S,2014ApJ...787...70M,2014A&A...571A..70F}. \citet{2017ApJ...845..110B} proposed that metallicity and/or magnetic activity effects are possible culprits for this radius inflation of TRAPPIST-1. However, without published evolutionary models that account for these effects, the authors were unable to validate this hypothesis. 

We present in this paper updated stellar parameters for TRAPPIST-1. We present a new parallax estimate in Sect. \ref{sect2}, allowing us to derive a more precise stellar luminosity. We next derive an updated stellar mass estimate for TRAPPIST-1, based on two approaches: an empirical mass derived from dynamical masses of equivalently classified ultracool dwarfs in astrometric binaries (Sect. \ref{sect3}), and a stellar mass from evolution modeling that is able to take into account metallicity and magnetic activity effects (Sect. \ref{sect4}). We combine the information from evolutionary models and dynamical masses in Sect. \ref{sect5} to obtain final stellar parameters for TRAPPIST-1. We conclude in Sect. \ref{sect6}. 

\section{New parallax and luminosity estimates}
\label{sect2}
\begin{figure}[!ht]
\begin{center}
\begin{tabular}{lll}
\includegraphics[scale=0.55,angle=0]{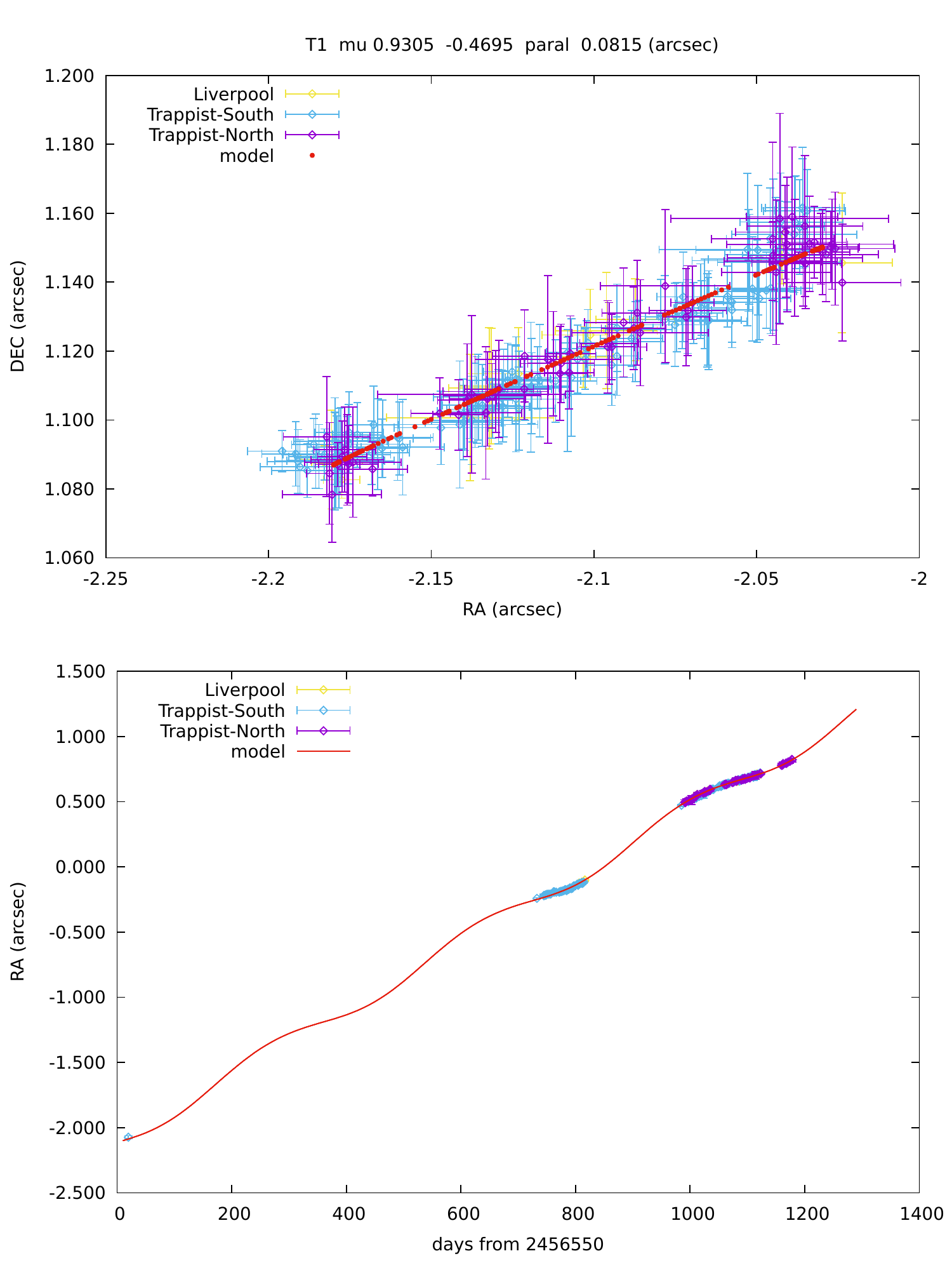}
\end{tabular}
\end{center}
\caption{\label{p1}\textit{Top}: RA-DEC parallactic displacement after subtraction of the proper motion (arbitrary origin). \textit{Bottom}: RA motion of TRAPPIST-1 as a function of the date (in days).}  
\end{figure}

\begin{figure}[!ht]
\begin{center}
\begin{tabular}{lll}
\includegraphics[scale=0.55,angle=0]{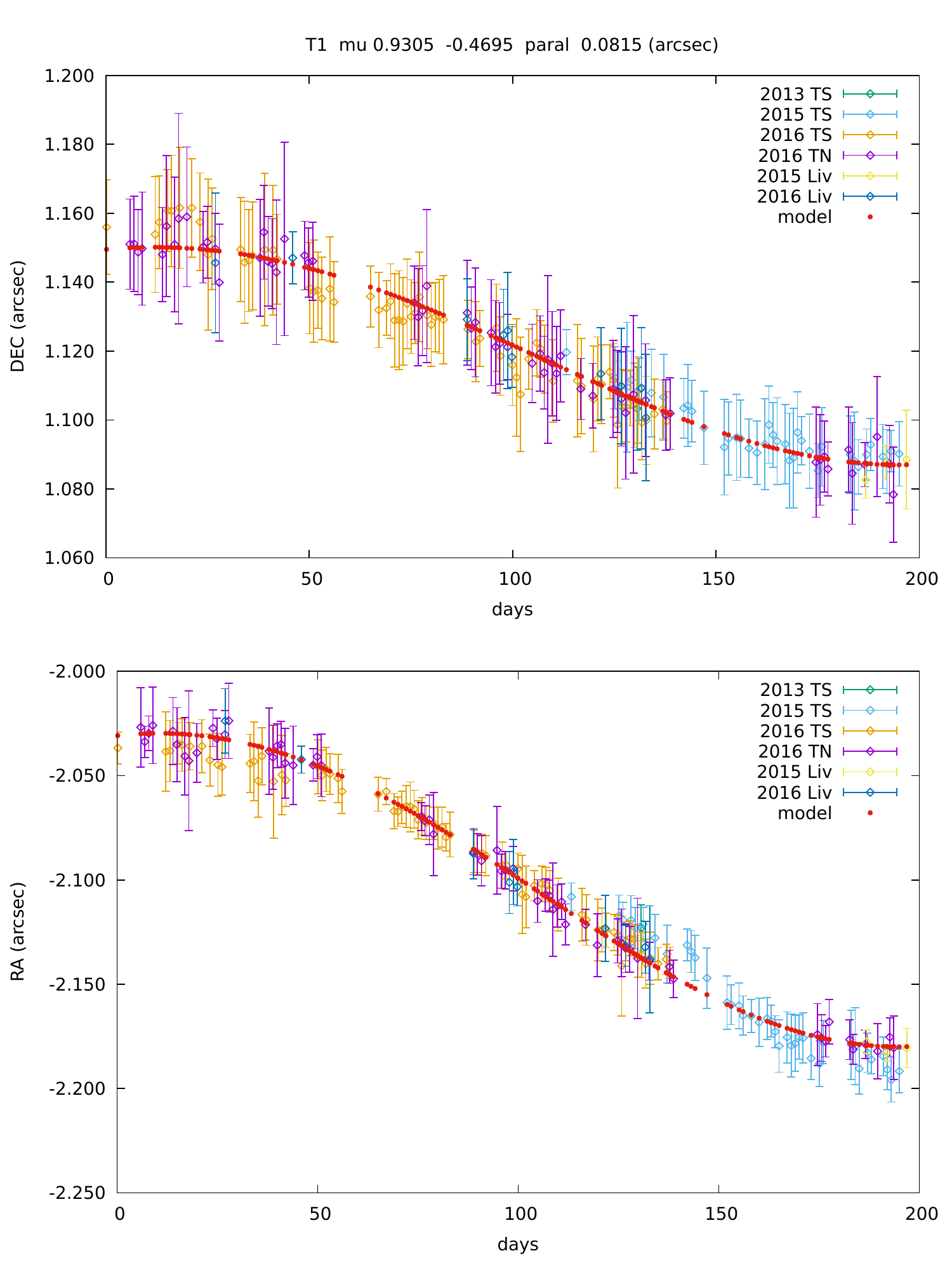}
\end{tabular}
\end{center}
\caption{\label{p2}\textit{Top}: DEC parallactic displacement as a function of the date, modulo 1 year (arbitrary origin). \textit{Bottom}: same, for RA.
}  
\end{figure}

In order to improve the distance and then luminosity measurements of TRAPPIST-1 we analysed all the optical data collected during the monitoring of TRAPPIST-1 to obtain its parallax as precisely as possible. Most data are from the UCDTS survey carried out with TRAPPIST-South (TS) located in the La Silla Observatory in Chile \citep{2011Msngr.145....2J,2011EPJWC..1106002G}. This data set consists of 33,118 images distributed among a total of 114 epochs regularly collected in 2013, 2015 and 2016 from May to December. They have been complemented with 10,969 images obtained in 2016 on 61 epochs with TRAPPIST-North (TN) located at the Oukaimeden Observatory in Morocco and 3,302 images on 13 epochs in 2015 and 2016 obtained with the Liverpool Telescope (LT; \citealt{2004SPIE.5489..679S}) in La Palma. The TS and TN images have been taken with an I+z filter and 2048x2048 pixels CCD cameras of 15 and 13 microns providing plate scales of 0.655$\arcsec$ and 0.60$\arcsec$ per pixel and covering fields of view of 22$\arcmin$ and 20$\arcmin$ respectively. The LT data have been taken using the IO:O camera and a Sloan z' filter and have a plate scale of 0.30$\arcsec$ per 2x2 binned pixel and a 10$\arcmin$ field of view. Each data set has been reduced separately using standard bias, dark and flat field correction techniques. Each epoch is a clear night for which there are at least 20 images spanning an hour angle of maximum 2.5 hours and for which we have obtained an astrometric solution with a rms $<$ 0.08", after keeping always the same best 65 stars having coordinates in the 2MASS catalogue. The model to fit the data is simply a constant proper motion and the periodic oscillation of the parallax. The best fit of all the data together gives a relative parallax of 0.0815$\arcsec$ and RA and DEC proper motions of 0.9305$\arcsec$ and $-$0.4695$\arcsec$ respectively (Fig. \ref{p1} and Fig. \ref{p2}). The fit to the individual data sets gives for the parallax and the proper motions (0.081$\arcsec$; 0.931$\arcsec$; $-$0.473$\arcsec$) for TS, (0.082$\arcsec$; 0.931$\arcsec$; $-$0.469$\arcsec$) for TN and (0.083$\arcsec$; 0.934$\arcsec$; $-$0.469$\arcsec$) for LT, respectively. This is in good agreement with the fit to the whole data set and allows us to also provide error bars from a weighted mean of the three data sets, giving for TRAPPIST-1 RAC and DEC proper motions of 0.9305 $\pm$ 0.0005$\arcsec$ and $-$0.4695 $\pm$0.0005$\arcsec$ respectively, and a relative parallax of 0.0815 $\pm$ 0.0006$\arcsec$. 

For the offset correction of the background reference stars to apply in order to obtain the absolute parallax, we used the mean value of Costa et al. 2006 (0.68 mas) and \citealt{2017AJ....154..103B} (1.08 mas), or 0.88 $\pm$ 0.20 mas. They both have fields of view similar to ours, use near IR filters like us, and have used stars in common to ours.

We then get a final value of the absolute parallax of 82.4 $\pm$ 0.8 mas (distance $d= 12.14 \pm 0.12$ pc). This measurement is in very good agreement with the parallax of 82.58 $\pm$ 2.7 mas provided by Costa et al. (2006), computed from 8 epochs and spanning 3 years. Our value is in disagreement with the more recent value of 79.59 $\pm$ 0.78 mas published by Boss et al (2017), based mostly on the data from \citet{2016AJ....152...24W}, and using 15 epochs from 2011 to 2016. In the latter study they noticed their parallaxes are about 2.5 mas smaller than those of stars in common with several other studies, which is about the difference we also find for TRAPPIST-1. An exquisite and definitive parallax value for TRAPPIST-1 should be obtained in coming years from Gaia data, with expected error as low as 0.05 mas, ten times better at least than our value.

Using our final value of the absolute parallax and the spectral energy distribution of \citet{2015ApJ...810..158F}, we revised the luminosity of TRAPPIST-1 to $L_*=(5.22\pm0.19)\times 10^{-4} L_{\odot}$, almost twice better than the previous uncertainties from \citet{2015ApJ...810..158F}.

\section{Empirical mass from astrometric binaries}
\label{sect3}
We made a first determination of the mass of TRAPPIST-1 by comparing to the dynamical masses inferred for a sample of ultracool dwarfs in astrometric binaries (\citealt{2017ApJS..231...15D}; hereafter DL17). We selected 20 objects from the sample of DL17 having a spectral type between M6 and L1.5, close to the one of TRAPPIST-1. We then used the NIR colors, luminosity, and mass of these 20 objects, and their associated errors from DL17 to estimate the mass of TRAPPIST-1 through the following Monte-Carlo approach. We performed an analysis composed of 100,000 independent steps. At each step, we drew a value for the luminosity and $J-K$ color index of TRAPPIST-1, from the normal distributions centered on 5.22$\times 10^{-4}$ and 1.058 with standard deviations of $1.9\times10^{-5}$ and 0.001, respectively. At each step, we also did the same for the 20 selected ultracool objects, drawing from the normal distributions corresponding to the values and errors from DL17. For each object $i$, the agreement between its luminosity and its $J-K$ index and those of TRAPPIST-1 was estimated - for both parameters - with the following formula: 

\begin{equation} 
\label{eqn1}
\delta x = \frac{|x_{T1}- x_i|}{\sqrt{(\sigma_{x_{T1}}^2 + \sigma_{x_i}^2)}}
\end{equation} 

where $x$ is $L_*$ or $J-K$ index, and $\sigma_{x}$ is its associated error, for TRAPPIST-1 (T1) or the object $i$. If $\delta x$ was larger than 1 for one of the two parameters, the object was discarded. For the remaining objects, a value of the mass was then drawn from the normal distribution corresponding to their mass measurement and error from DL17, and stored. At the end of the 100,000 steps, we obtained a resulting mass PDF with mean and standard deviation of $M_*=0.090 \pm 0.012 M_\odot$. Exactly same results were obtained using $H-K$ or $J-H$ color indexes. 
\section{Stellar mass from evolution modeling}
\label{sect4}
\subsection{Evolutionary models for very low-mass stars}
We adapted our stellar evolution code CLES (Code Li\'egeois d'Evolution Stellaire) to compute structures of very low-mass stars (VLMS). We refer to \citet{2008Ap&SS.316...83S} for the main constitutive physics and numerical features, but here are the details specific to VLMS. Two aspects are of particular relevance for computing structures of VLMS \citep{1997A&A...327.1039C}: the surface boundary conditions, which must be extracted from detailed model atmospheres, and the equation of state (EOS), which must cover the dense and cool regime of  VLMS. We extracted boundary conditions from the publicly available BT-Settl model atmospheres \citep{2012RSPTA.370.2765A,2012EAS....57....3A,2013A&A...556A..15R} computed with the solar abundances of \citet{2009ARA&A..47..481A}. Several compositions are publicly available, from $Z_{\rm ini}\sim0.004$ to $\sim0.04$. We also used the solar abundances of \citet{2009ARA&A..47..481A} to compute the interior structure. The transition interior/atmosphere is performed at an optical depth $\tau=100$, similarly to the stellar evolution models of BHAC15 that are the commonly used reference for VLMS. For the EOS, we considered H, He, C, and O. We directly adapted tables built for white dwarfs and subdwarf B stars kindly made available to us by G. Fontaine. These tables cover a large domain of the temperature-density plane including VLMS (details are given in \citealt{2013A&A...553A..97V}). In a nutshell, in the partial ionization region where nonideal and degeneracy effects are important, we used the EOS of \citet{1995ApJS...99..713S} for H and He, an improved version of the EOS of \citet{1977ApJS...35..293F} for C, and similar developments for the EOS of O. Interpolation in composition is handled following the additive volume prescription of \citet{1977ApJS...35..293F}. We used opacities from the OPAL project \citep{1996ApJ...464..943I}, combined for low temperatures to opacities from \citet{2005ApJ...623..585F}. The effects of thermal conductivity have been taken into account according to the computations of \citet{1999A&A...346..345P} and \citet{2007ApJ...661.1094C}. Nuclear reaction rates, for De and Li burning, as well as for the pp chain, come from the NACRE II compilation \citep{2013NuPhA.918...61X}. Convection is treated using the mixing length theory (MLT). For ultracool stars, we set $\alpha_{\rm MLT}$ (the ratio between the mixing length and the pressure scale height) to 2.0, according to recent 3D radiative hydrodynamic (RHD) simulations, showing that the calibrated $\alpha_{\rm MLT}$ increases to this value for the coolest and densest stars \citep{2015A&A...573A..89M}. Our solar calibration (evolutionary track giving the Sun at the right age, luminosity and effective temperature, here without diffusion) gives $\alpha_{\rm MLT}=1.8$, $X_{\rm ini}=0.728$, and $Z_{\rm ini}=0.013$. Our initial helium abundance is therefore close to initial helium abundance of Bt-Settl model atmospheres, which is $Y_{\rm ini}=0.249$ (with \citealt{2009ARA&A..47..481A} solar mixture). BHAC15 however showed the exact consistency of $Y_{\rm ini}$ between interior structure and model atmosphere is not a source of tension, especially for VLMS (see their Sect. 4.2).

\begin{figure}[!ht]
\begin{center}
\begin{tabular}{lll}
\includegraphics[scale=0.35,angle=0]{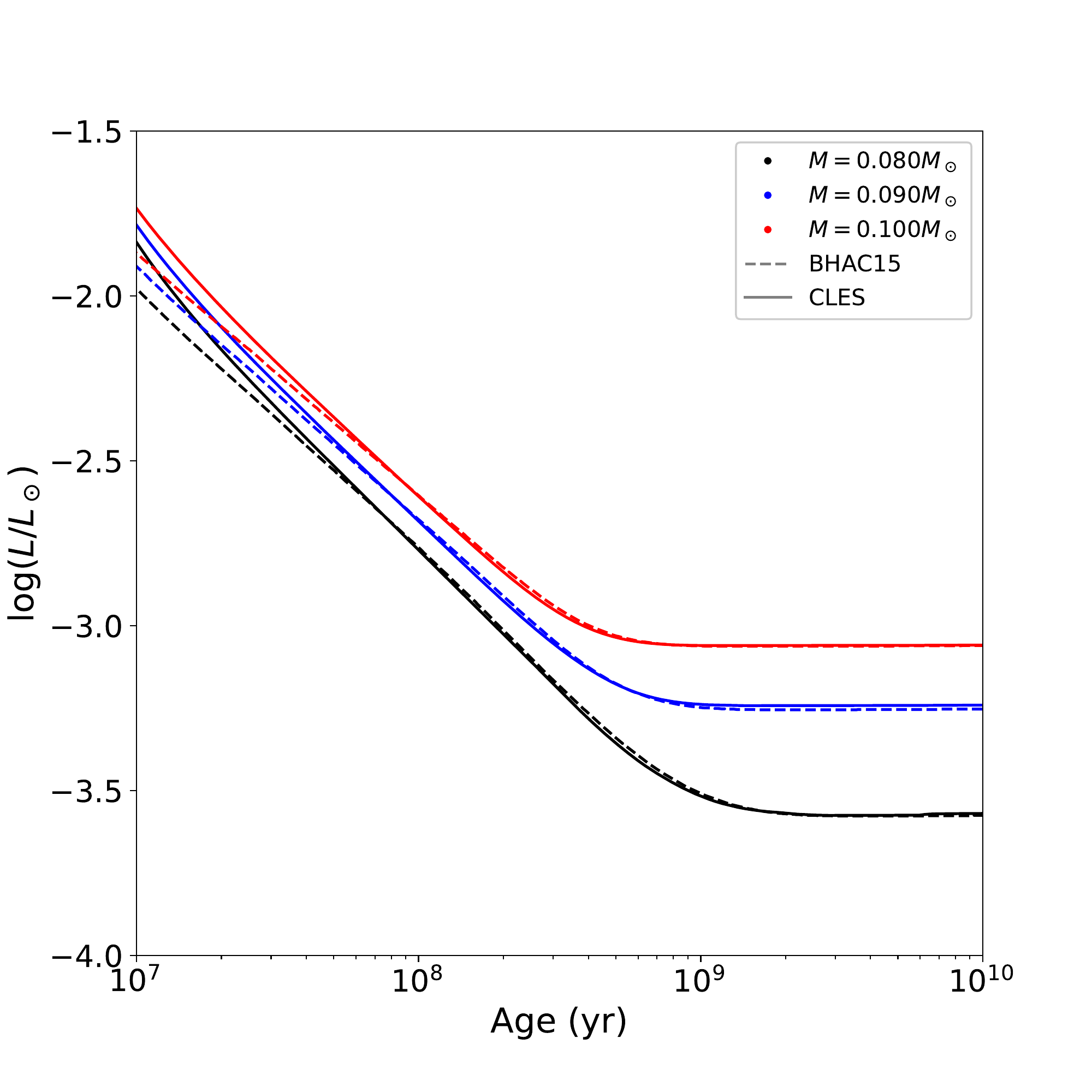}\\
\includegraphics[scale=0.35,angle=0]{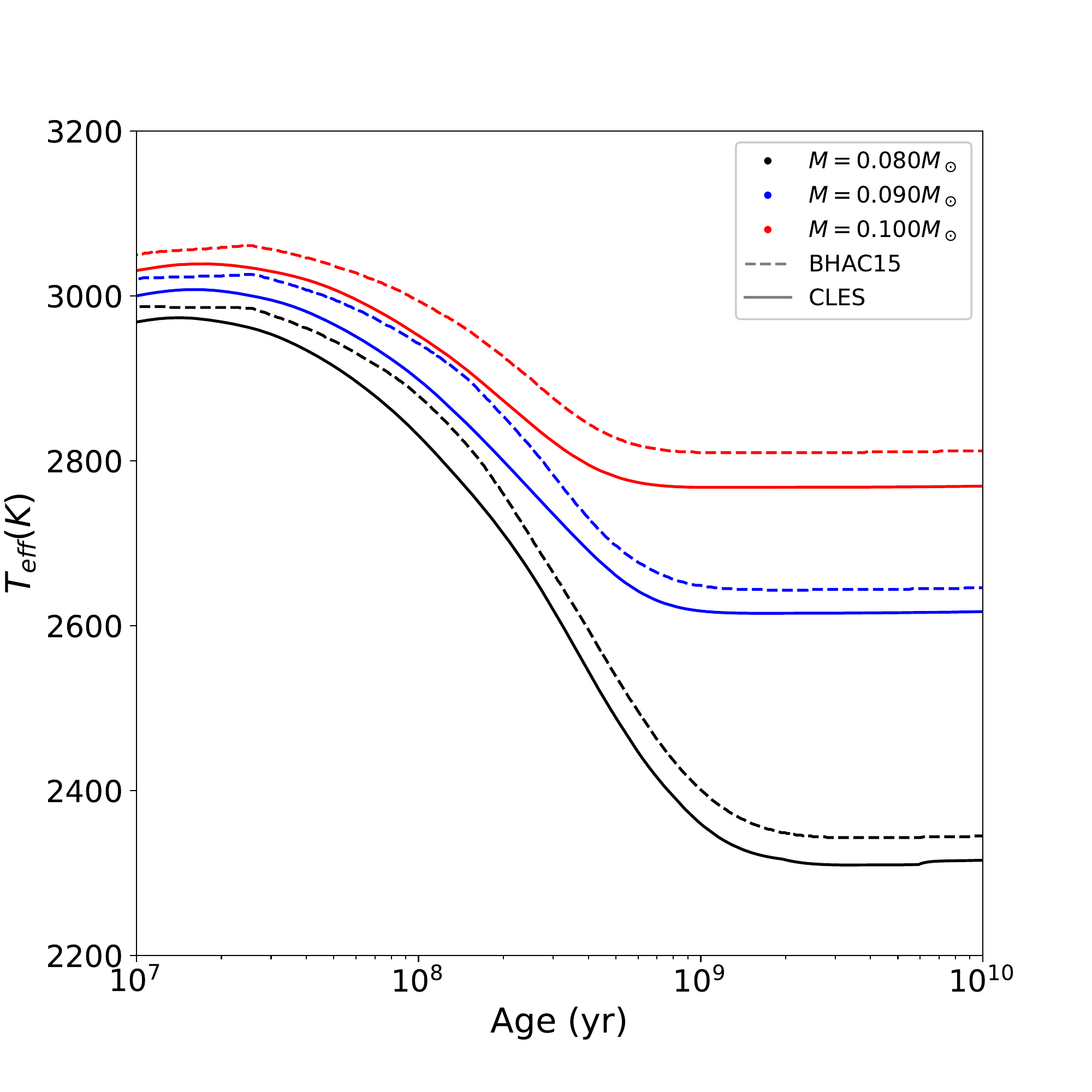}\\
\includegraphics[scale=0.35,angle=0]{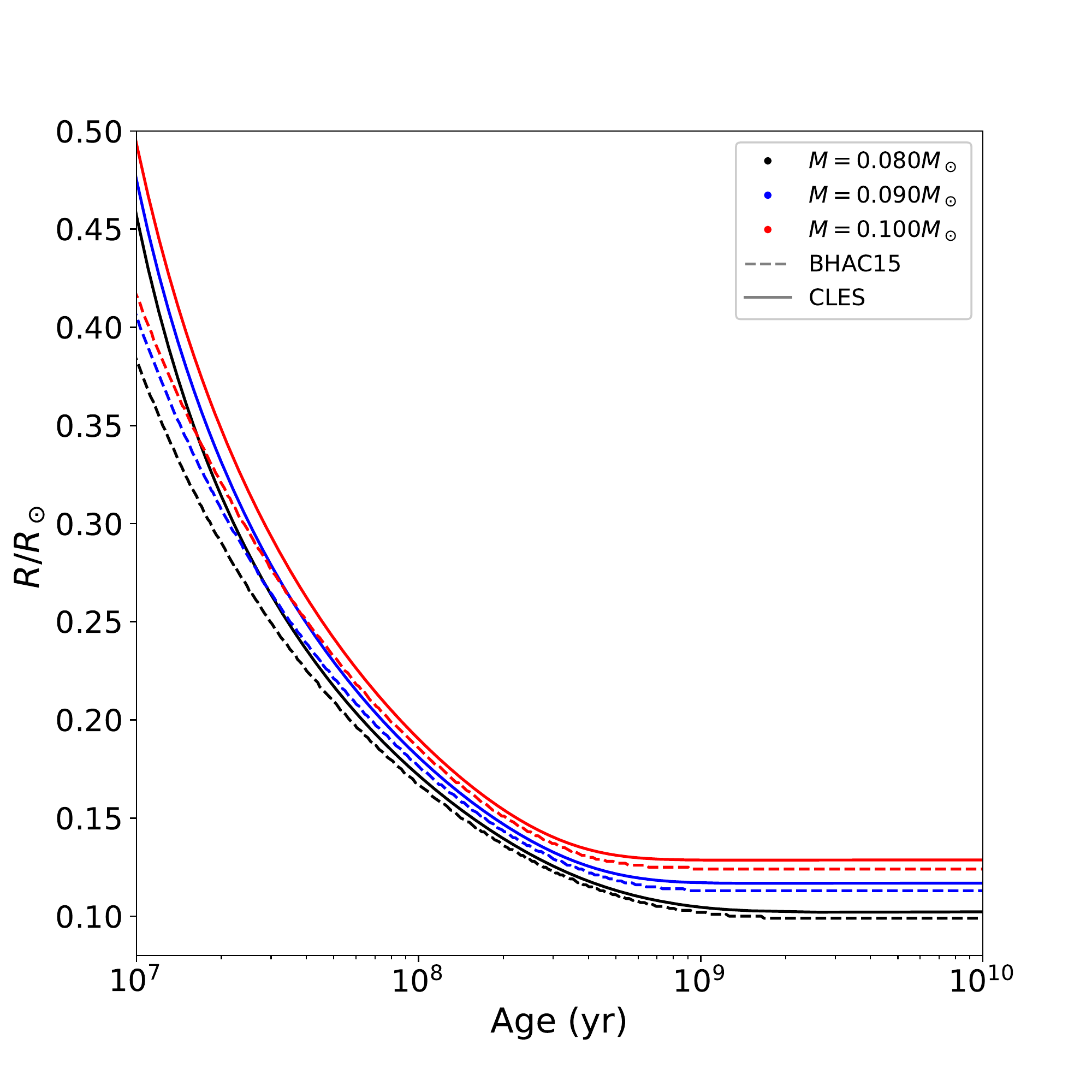}
\end{tabular}
\end{center}
\caption{\label{f1}Comparison between CLES and BHAC models for very similar input physics, for luminosity (top panel), effective temperature (middle panel), and radius (bottom panel), for 0.08, 0.09 and 0.10 $M_{\odot}$.}  
\end{figure}

A comparison between CLES and BHAC15 models is provided on Fig. \ref{f1} that shows evolutionary tracks for 0.08, 0.09 and 0.10 $M_{\odot}$ star with, as far as possible, identical constitutive physics  (\citealt{1993oee..conf...15G} for the interior/\citealt{2009ARA&A..47..481A} supplemented by \citealt{2011SoPh..268..255C} for some elements for the boundary conditions; EOS for H and He only; Chabrier \& Baraffe 1997; BHAC15). BHAC15 and CLES stellar tracks are very close. CLES models tend to provide, for a given mass, very similar luminosity with slightly larger stellar radius (by $\sim$ +3\%) -and therefore lower stellar density- and slightly lower effective temperature (by $\sim$ $-$30 K) estimates compared to BHAC15 models. Our CLES models therefore have similar strengths and weaknesses than BHAC15 models: they tend to provide accurate estimates for the mass and the luminosity, while they tend to underestimate the radius and overestimate the effective temperature \citep[e.g.][and references therein]{2005nlds.book.....R,2013AN....334....4T,2013ApJ...776...87S,2017ApJS..231...15D}. This is however not a systematic trend, and some stars are consistent with theoretical estimates \citep{2013ApJ...776...87S,2016A&A...593A.127K,2017A&A...604L...6V}.

\begin{figure}[!ht]
\begin{center}
\begin{tabular}{lll}
\includegraphics[scale=0.35,angle=0]{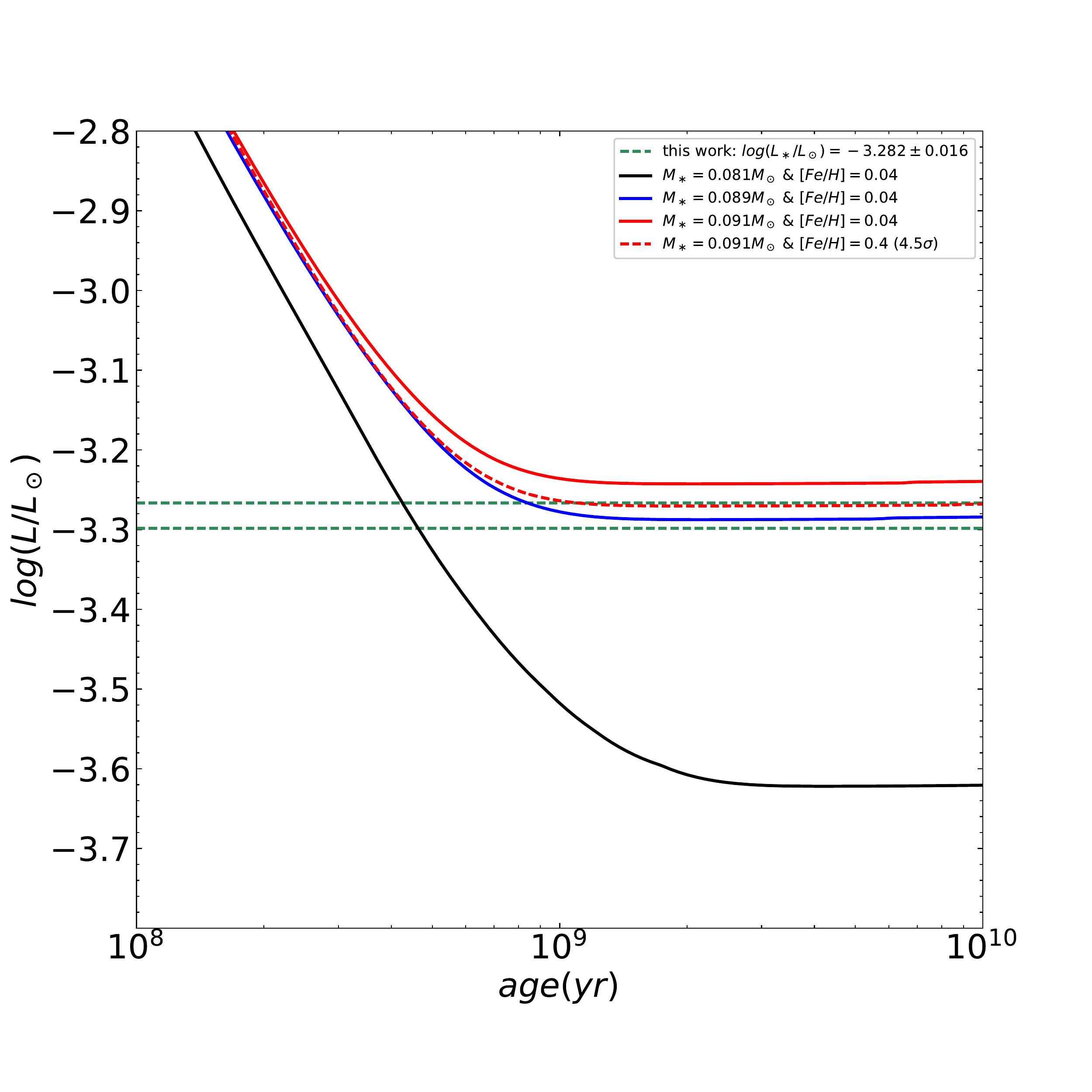}\\
\includegraphics[scale=0.45,angle=0]{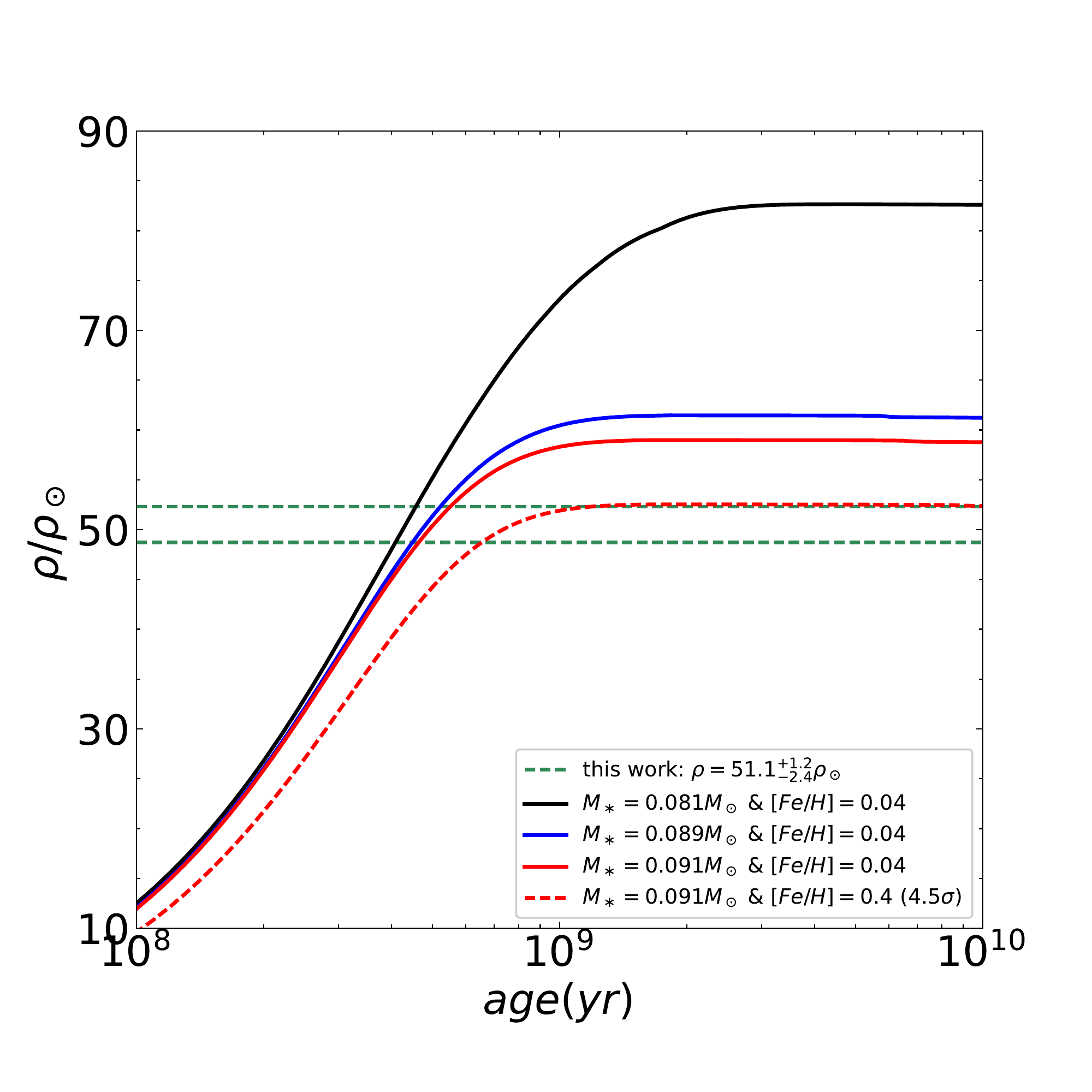}
\end{tabular}
\end{center}
\caption{\label{f2}Stellar luminosity (top panel) and density (bottom panel) for evolution models for various masses and metallicities, compared to luminosity and density estimates for TRAPPIST-1 (dashed horizontal lines).}  
\end{figure}

\subsection{Modeling TRAPPIST-1}
We performed stellar evolution modeling for TRAPPIST-1 using our in-house Levenberg-Marquardt optimization algorithm \citep{1992nrfa.book.....P}. Three independent measurements were used to constrain stellar models: luminosity (see Sect. \ref{sect2}), age \citep{2017ApJ...845..110B}, and density inferred from transits. We took the most recent value inferred for the most complete observational data set (Delrez et al., submitted): $\rho_*= 51.1^{+1.2}_{-2.4}  \rho_{\odot} $. The iron abundance of TRAPPIST-1, [Fe/H]$=0.04\pm0.08$, corresponds, assuming all elements scale like in the Sun, to a $Z/X$ ratio of the models of 0.020$\pm$0.004. In all stellar evolution modeling, the reference values are given for $Y_{\rm ini}=0.26$, the initial helium abundance of the Sun according to our solar calibration.

\subsubsection{First case: luminosity and density as constraints, no age indication}
\label{first}
Firstly, we checked with our CLES models the stellar estimates used so far as priors in transit analyses (see Introduction and \citealt{2016Natur.533..221G,2017Natur.542..456G}). No indication on the age of TRAPPIST-1 was available at that time. Using stellar density and luminosity as constraints, we obtained a stellar mass $M_*=0.081\pm 0.003 M_{\odot}$ and age $=450 \pm 55$ Myr, which corresponds to model radius $R_*=0.117 \pm 0.002 R_{\odot}$ and model effective temperature $T_{\rm eff} = 2555 \pm 25$ K. These errors come from errors $L_*$, $\rho_*$ and $Z/X$, based on error propagation with full evolutionary tracks. More precisely, we computed various evolutionary tracks by varying observational constraints ($L_*$, $\rho_*$ and $Z/X$) within their given 1-$\sigma$ range and computed the respective 1-$\sigma$ confidence interval for $M_*$, $R_*$, $T_{\rm eff}$ and age.  

The parameters we derived with CLES models are in complete agreement with those used so far for transit analyses (see Introduction and \citealt{2016Natur.533..221G,2017Natur.542..456G}). As previously noticed, this corresponds to a young age for TRAPPIST-1, which is now disputed \citep{2017ApJ...845..110B}. The priors on stellar parameters must then be revised.

\subsubsection{Second case: luminosity and age as constraints}
Secondly, we performed stellar evolution modeling using luminosity and age as constraints only, given the discrepancies of the models towards radius and effective temperature. It can directly be seen on Fig. \ref{f2} (top panel) that a stellar mass of $\sim$0.09 $M_{\odot}$ is needed to account for the old age and luminosity of TRAPPIST-1. More quantitatively, we found a stellar mass of $M_*=0.089 \pm 0.003 M_{\odot}$ for an age between 2-15 Gyr (evolution models are not able to provide a precise stellar age, as the star evolves extremely slowly). This error was computed as in Sect. \ref{first}, and it took also into account the unknown initial helium abundance, varied from $Y=0.25$ (primordial value) to $Y=0.30$, a reasonable assumption for a field star like TRAPPIST-1 (see, for instance, \citealt{2014ApJS..214...27M} for the initial helium abundance of 42 \textit{Kepler} stars inferred from asteroseismology). These errors have been quadratically added to the previous ones. The corresponding model radius is $R_*=0.114 \pm 0.002 R_{\odot}$ and effective temperature is $T_{\rm eff} = 2595 \pm 30$ K. Let us note here that systematic errors of stellar models are notoriously difficult to estimate. We varied the depth of the transition between interior and atmosphere (from the reference at $\tau=100$ up to the photosphere), as well as the $\alpha_{\rm MLT}$ parameter of convection (from solar to the reference value) and the opacities (using Opacity Project (OP) table rather than OPAL; \citealt{2005MNRAS.360..458B}). No significant difference on the results have been found. Other constitutive physics cannot be easily varied (in particular, no other EOS is currently available for the high-density, low temperature domain encompassed by compact objects with masses below 0.1 $M_{\odot}$). 

\subsubsection{Third case: luminosity, density and age as constraints}
Finally, we used luminosity, age and density as constraints for stellar evolution modeling. At solar metallicity, no reasonable fit was found. Indeed, it can directly be seen on Fig. \ref{f2} that the stellar density at $\sim$0.09$M_{\odot}$ would be much higher than the value measured from transits, due to a too low stellar radius. As already noticed, the CLES models do not provide a better job than the BHAC15 models related to the radius (and then stellar density) discrepancies. The usual suspects for this radius anomaly are the presence of strong magnetic field and/or magnetic activity like spots, causing the stars to inflate by inhibiting convective energy transport  \citep{2001ApJ...559..353M,2007A&A...472L..17C}, or increased metallicity compare to solar, causing the star to inflate by increased stellar material opacity \citep{2014A&A...571A..70F}. 

We empirically found that by doubling the $Z/X$ ratio (corresponding to [Fe/H]=0.40, a +4.5$\sigma$ error on the available estimate), we were able to reconcile stellar density and luminosity with the old age of \citet{2017ApJ...845..110B}. By performing a new Levenberg-Marquardt optimization, we found a stellar mass of $M_*=0.091 \pm 0.005 M_{\odot}$ for an age between 2-15 Gyr. The corresponding model radius is $R_*=0.120 \pm 0.002 R_{\odot}$ and effective temperature is $T_{\rm eff} = 2530 \pm 35$ K. The quoted errors are computed as previously. 
 
Very similar results with a good fit to luminosity and density at old age were obtained by greatly reducing convection efficiency (down to $\alpha_{\rm MLT}\sim 0.05$). The two usual suspects for radius inflation, metallicity and magnetic activity effects, are therefore possible to account for the stellar density of TRAPPIST-1. TRAPPIST-1 lies at the transition between thin and thick disk (\citealt{2015ApJS..220...18B}, \citealt{2017ApJ...845..110B}). It is possible that the [Fe/H] measurement obtained by near-infrared spectroscopy \citep{2016Natur.533..221G} is biased towards lower values by C and O abundances, which affect the pseudo-continuum level \citep{2016ApJ...828...95V}. Investigations on high-resolution spectra to identify $\alpha$-elements abundances and hence, to determine TRAPPIST-1 metallicity is a work to be done. Although rare, supermetallic stars exist in the solar neighborhood, such as Alpha Cen \citep{2008A&A...488..653P,2005A&A...441..615M}. On the other hand, a convection parameter $\alpha_{\rm MLT}\sim 0.05$ is a huge reduction of the convection efficiency. \citet{2014ApJ...789...53F} demonstrated that magnetic stellar models are indeed unable to significantly inflate fully convective stars, unless are present extremely strong interior magnetic fields, and/or high-coverage, clustered at poles star spots. TRAPPIST-1 is a low-activity M8 star \citep{2017NatAs...1E.129L} with a moderate surface magnetic field of 600$^{+200}_{-400}$ G \citep{2010ApJ...710..924R}, for which an interior magnetic field of several MG (necessary to significantly inflate the star) is difficult to imagine. Brightness inhomogeneities are indeed present on TRAPPIST-1 \citep{2017NatAs...1E.129L}, but a full analysis to determine their coverage and repartition over the star has still to be done to confirm or refute this hypothesis.

Alternatively, such enhanced metallicity or convection reduction may actually correspond to missing/perfectible constitutive physics in stellar models, related to opacities, model atmospheres or EOS. We will investigate these possibilities in forthcoming papers.

\begin{figure}[!ht]
\begin{center}
\begin{tabular}{lll}
\includegraphics[scale=0.35,angle=0]{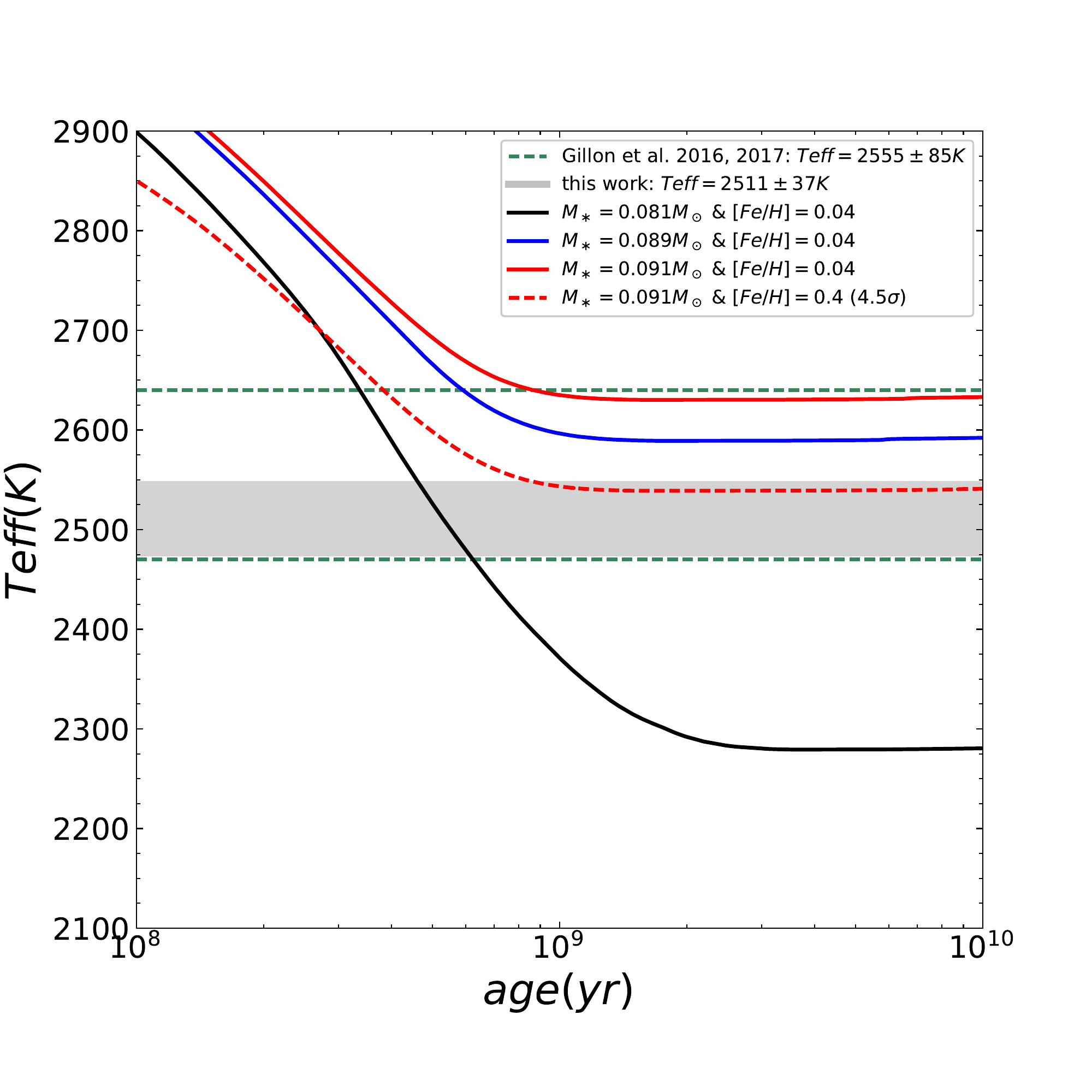}\\
\includegraphics[scale=0.35,angle=0]{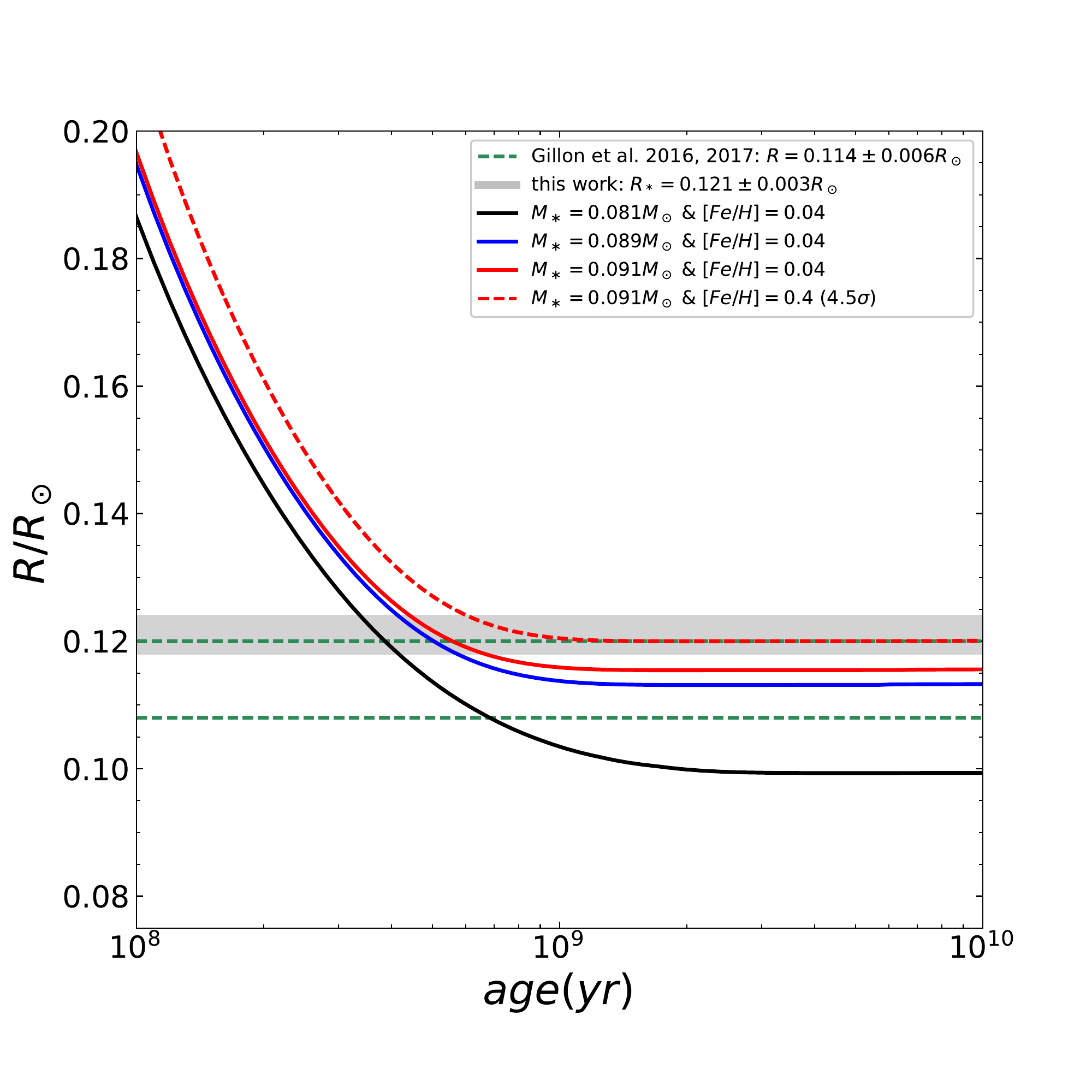}
\end{tabular}
\end{center}
\caption{\label{f3}Effective temperature (top panel) and stellar radius (bottom panel) for evolution models for various masses and metallicities. Horizontal dashed lines shows the previous prior $T_{\rm eff}$ and $R_*$ PDFs used for deriving planetary parameters \citep{2017Natur.542..456G} and grey areas represent updated estimates from this work.}  
\end{figure}

\section{Final stellar parameters for TRAPPIST-1}
\label{sect5}
Finally, we combined the information from stellar evolution models and ultracool dwarf binaries to obtain final stellar parameters for TRAPPIST-1. Given the discrepancies of stellar models towards radius and effective temperature, we relied on the stellar mass obtained by using luminosity and age as constraints only (sect. 4.2.2). We carried out the same Monte-Carlo analysis described in Sect. \ref{sect3}, except for the following. At each step, for each tested ultracool object from DL17, the value drawn for its mass was compared to a value drawn from the distribution $N(0.0 89,0.003^2)$, using again equation (\ref{eqn1}) to discard values disagreeing with each other at more than 1-$\sigma$. We obtained $M_*=0.089 \pm 0.006 M_{\odot}$.

At each step of the analysis we also drew values for the density and luminosity of TRAPPIST-1 based on the measurements $\rho_*= 51.1^{+1.2}_{-2.4}  \rho_{\odot} $(Delrez et al. submitted) and  $L_*=(5.22\pm0.19)\times 10^{-4} L_{\odot}$ (Sect. \ref{sect2}), enabling to compute a value for the stellar radius (from $\rho_*$ and $M_*$) and for the effective temperature (from $L_* = 4\pi R_\ast^2 \sigma T_{\rm eff}^4$). The means and standard deviations of the distributions resulting from the 100,000 steps are $R_* = 0.121 \pm 0.003 R_\odot$ and $T_{\rm eff}$ = 2516 $\pm$ 41 K. We adopt these values and errors as our updated stellar parameters for TRAPPIST-1, which will be used in our forthcoming transit analyses (e.g. Delrez et al. submitted). These values are summarized in Table 1.

\begin{table}[!ht]
\label{blabla}
\caption{Updated stellar parameters of TRAPPIST-1.}
\begin{center}
\begin{tabular}{ll}
\hline \hline
Quantity  & Value\\
\hline
\hline
$L_*/L_{\odot}$  & 0.000522$\pm$ 0.000019 \\
$M_*/M_{\odot}$  & 0.089$\pm$ 0.006 \\
$R_*/R_{\odot}$ $^1$  &  0.121 $\pm$ 0.003 \\
$T_{\rm eff}$ (K) $^2$  & 2516 $\pm$ 41\\
\hline
\multicolumn{2}{l}{
{\footnotesize $^1$ From $M_*$ and $\rho_*$}}\\
\multicolumn{2}{l}{
{\footnotesize $^2$ From $L_*$ and $R_*$}
}
\end{tabular}
\end{center}
\end{table}
\section{Conclusions}
\label{sect6}
We presented in this paper updated estimates for the TRAPPIST-1 star. We proposed a new measurement of its parallax, 82.4 $\pm$ 0.8 mas, based on 188 epochs and 45,000 images with the TRAPPIST and Liverpool telescopes. This lead to a revised luminosity of $L_*=(5.22\pm0.19)\times 10^{-4} L_{\odot}$, almost twice more precise than the previous estimate. We also proposed an updated mass based on two independent approaches, stellar evolution modeling and an empirical model-independent methodology based on astrometric binaries. We combined this information to obtain the final stellar mass for TRAPPIST-1: $M_*=0.089 \pm 0.006 M_{\odot}$. Combined to stellar density from transits, this mass lead to $R_* = 0.121 \pm 0.003 R_{\odot}$ which, combined to luminosity, gave $T_{\rm eff} =$ 2516 $\pm$ 41 K.


The stellar parameters we propose in this paper represents a significant improvement compared to the priors used in previous transit analyses  \citep{2016Natur.533..221G,2017Natur.542..456G}, which were based on stellar evolution models only and corresponds to a young star, which is discarded for TRAPPIST-1. 
The exact impact on planets properties, particularly on their masses inferred from TTVs (hence on planetary densities), on their irradiation (hence on surface conditions and habitability), and on their atmospheric evolution (the contraction time onto the main sequence is $\sim$1.9 Gyr for a 0.09$M_\odot$, compared to $\sim$5.8 Gyr for a 0.08$M_\odot$) will be assessed in future studies (e.g. Delrez et al., submitted; Grimm et al. submitted).



\begin{acknowledgements}
We thank Trent Dupuy for discussions of VLMS binary masses. We are indebted to Gilles Fontaine and Pierre Brassard for having provided us EOS tables. We warmly thank the stellar physics team in Liege for their advices when developing ultracool stellar models. V.V.G. and M.G. are F.R.S.-FNRS Research Associates. E.J. is F.R.S-FNRS Senior Research Associate. C.S.F. is funded by an Action de Recherche Concert\'ee (ARC) grant financed by the Wallonia-Brussels Federation. TRAPPIST-South is a project funded by the Belgian Fonds (National) de la Recherche Scientifique (F.R.S.-FNRS) under grant FRFC 2.5.594.09.F, with the participation of the Swiss National Science Foundation (FNS/SNSF). TRAPPIST-North is a project funded by the University of Li\`ege, and performed in collaboration with Cadi Ayyad University of Marrakesh. The research leading to these results has received funding from the European Research Council under the FP/2007-2013 ERC Grant Agreement number 336480, and from the ARC grant for Concerted Research Actions, financed by the Wallonia-Brussels Federation. Our paper also uses data obtained at the Liverpool Telescope, which is operated on the island of La Palma by Liverpool John Moores University in the Spanish Observatorio del Roque de los Muchachos of the Instituto de Astrofisica de Canarias with financial support from the UK Science and Technology Facilities Council. This work was partially supported by a grant from the Simons Foundation (PI Queloz, grant number 327127). L.D. acknowledges support from the Gruber Foundation Fellowship. A.J.B. acknowledges support from the US-UK Fulbright Commission. B.-O.D. acknowledges support from the Swiss National Science Foundation (PP00P2-163967).

\end{acknowledgements}

\software{CLES (Scuflaire et al. 2008)}



\end{document}